\documentclass[twocolumn,showpacs,aps]{revtex4}

\usepackage{graphicx}
\usepackage{dcolumn}
\usepackage{bm}

\begin{document}

\title{Oxygen Isotope Effect of the Plane-Copper NQR Frequency in 
         $\mathbf{YBa_2Cu_4O_8}$} 
\author{M. Mali}
 \email{mali@physik.unizh.ch}
\author{J. Roos, and H. Keller} 
\affiliation{Physik--Institut, Universit\"{a}t Z\"{u}rich, CH--8057 Z\"{u}rich, 
         Switzerland} 
\author{J. Karpinski} 
\affiliation{Laboratorium f\"{u}r Festk\"{o}rperphysik, Eidgen\"{o}ssische 
         Technische Hochschule Z\"{u}rich, CH--8093 Z\"{u}rich, Switzerland}
\author{K. Conder} 
\affiliation{Laboratory for Neutron Scattering, ETH Z\"urich and PSI Villigen, 
CH--5232 Villigen PSI, Switzerland}

\date{\today}
             
\begin{abstract}
We report on high-precision measurements of the temperature dependence 
of the plane-$^{63}$Cu NQR line frequency $\nu_{Q}$(Cu2) and the 
linewidth in normal and superconducting $^{16}$O and $^{18}$O 
exchanged $\mathrm{YBa_2Cu_4O_8}$.  Whereas $\nu_{Q}$(Cu2) passes 
$T_{c}$ very smoothly without a discontinuity either in value nor in 
slope, the linewidth increases in the normal conducting phase down to 
$T_{c}$ and starts to decrease sharply in the superconducting phase to 
finally resume its high-temperature value of the normal phase.  \\
There is a well discernible oxygen isotope effect on the 
$\nu_{Q}$(Cu2) temperature dependencies.  The temperature dependence 
of $\nu_{Q}$(Cu2) is described by an empirical expression consisting 
of two parts: one related to the thermal expansion of the lattice and 
the other due to charge redistribution during the formation of new 
electronic structures in the CuO$_{2}$ planes.  From the fit to the 
experimental data we determine for the conjectured formation of new 
electronic structures an energy scale $\Delta$($^{16}$O) = 
188.0(1.6)~K and $\Delta$($^{18}$O) = 180.0(1.6)~K.  This results in a 
partial oxygen isotope effect coefficient $\alpha_{\nu_{Q}} = 
0.42(11)$ which is larger than both the spin-pseudogap coefficient 
$\alpha_{PG} = 0.061(8)$ and the $T_{c}$ coefficient $\alpha_{T_{c}} = 
0.056(12)$ \cite{Raffa}.
\end{abstract}
\pacs{74.72.Bk, 76.60.Gv}
\maketitle
\section{INTRODUCTION}\label{intro}
In the past few years very strong experimental evidence has been found 
for static or dynamic charge inhomogeneities in strongly correlated 
electronic systems in particular in high-$T_{c}$ superconductors where 
an array of self organized one dimensional structures known as stripes 
can appear and with them the possibility of a pseudogap \cite{Emery}.  
Static charge redistributions might occur at the formation of new 
electronic structures.  The nuclear quadrupole resonance (NQR) 
frequency ($\nu_{Q}$) being proportional to the electric field 
gradient (EFG) can serve as a very sensitive monitor of any such 
charge redistribution.  The plane-copper (Cu2) nuclear quadrupole 
resonance frequency $\nu_{Q}$(Cu2) in $\mathrm{YBa_2Cu_4O_8}$ seems to 
signal such an event.  Besides the effect of thermal lattice 
expansion, the temperature dependence of $\nu_{Q}$(Cu2) reveals an 
additional effect possibly connected to the charge redistribution in 
the CuO$_{2}$ planes induced by some kind of a new electronic 
structure.  The temperature dependence of $\nu_{Q}$(Cu2) is very 
unconventional, particularly striking is the change of the course 
around 190~K where a broad minimum in the frequency appears 
\cite{Zimmermann}.  In the same temperature region other anomalies in 
nuclear magnetic resonance (NMR) and NQR quantities, for instance in 
Knight shifts, line widths, and relaxation times occur which signal an 
electronic crossover \cite{Suter1,Suter2}.  To learn more about the 
CuO$_{2}$ planes especially about a possible charge redistribution 
during the pseudogap formation as well as to have very precise 
$\nu_{Q}$(Cu2) values frequently demanded in other Cu2 NMR and NQR 
studies we decided to implement the Cu2 $\nu_{Q}$ and linewidth data 
by additional measurements on the very same material on which we 
recently determined by Cu2 spin-lattice relaxation the oxygen isotope 
effect of the spin-pseudogap \cite{Raffa}.  Therefore we performed 
between 5~K and 350~K very accurate measurements of the temperature 
dependence of the Cu2 NQR line frequency $\nu_{Q}$(Cu2) and the 
linewidth in normal and superconducting $^{16}$O and $^{18}$O 
exchanged $\mathrm{YBa_2Cu_4O_8}$.  It is our anticipation that 
accurate knowledge of the influence of different oxygen isotopes onto 
$\nu_{Q}$(Cu2) and its temperature dependence will add clues to the 
understanding of high-$T_{c}$ superconductors.
\section{EXPERIMENTAL DETAILS}\label{exp}
Sample preparation, oxygen exchange process and sample 
characterization are described in Ref.~\cite{Raffa}.  The $^{18}$O 
content in the $^{18}$O sample is 88$\%$ as determined by the weight 
loss of $^{18}$O material after back exchange of $^{18}$O with 
$^{16}$O.  Room temperature x-ray measurements show a small difference 
in the lattice parameters of the two oxygen isotope samples.  The $a$, 
$b$, and $c$ lattice parameters are 3.8411(1)~\AA, 3.8717(1)~\AA, 
27.2372(8)~\AA ~for the $^{16}$O and 3.8408(1)~\AA, 3.8718(1)~\AA, 
27.2366(8)~\AA ~for the $^{18}$O samples.  For the detection of the 
NQR signal we used a standard NQR pulse spectrometer employing 
spin-echo technique and echo recording in quadrature.  The magnitude 
of the complex Fourier transform of the whole echo delivered the shape 
and the position of the line.  The two oxygen exchanged samples were 
inserted into a probe head with two identical resonance circuits which 
allowed simple switching of the electronics from one sample to the 
other thus minimizing the effects of any possible slow drifts in the 
characteristics of the equipment and temperature.  To guarantee a 
proper excitation and detection of the relatively broad line we 
increased the damping of the resonance circuits by additional 7 
$\Omega$ resistances.  Further we used 6 mm diameter coils and highest 
power available (1 kW) such that a pulse of only 1.8 $\mu$s duration 
corresponded to a $\pi$ pulse.  We also took care to tune the 
resonance circuits as well as to excite the line with pulses at a 
frequency very close to the middle of the line, the difference between 
the two never exceeding more than 10~kHz.  During the whole experiment 
we followed a rigid measurement procedure to ensure an equal treatment 
of the two samples.
\section{ANALYSIS AND RESULTS}\label{results}
\begin{figure}[h]
 	\centering
 	\includegraphics[width=\linewidth]{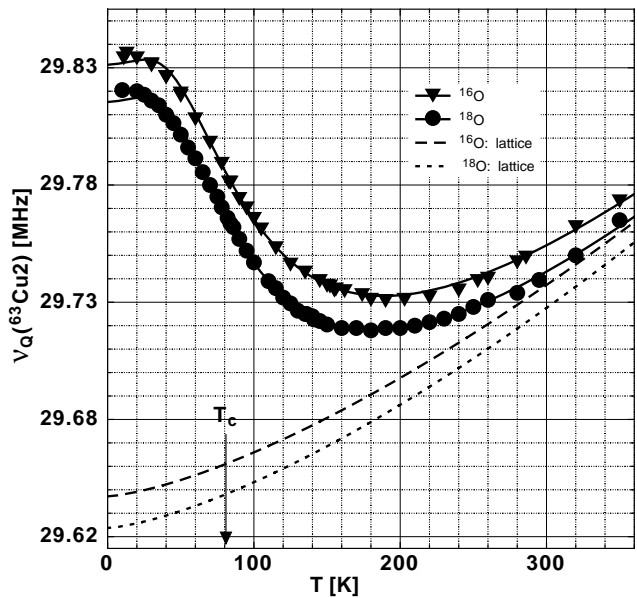} \vskip12pt
 	\caption{Temperature dependence of the NQR frequency of plane $^{63}$Cu2 in 
 	$^{16}$O (full triangles) and $^{18}$O (full circles) exchanged
 	 $\mathrm{YBa_2Cu_4O_8}$. The solid lines are fits of Eq.~(\ref{nuQofT}) 
 	 to the experimental data and the broken lines represent the conjectured 
 	 temperature dependence of $\nu_{Q}$($^{63}$Cu2) 
 	 that comes from the thermal expansion of the lattice.}  
 	\label{freq}
\end{figure}

\begin{figure}[h]
 	\centering
 	\includegraphics[width=\linewidth]{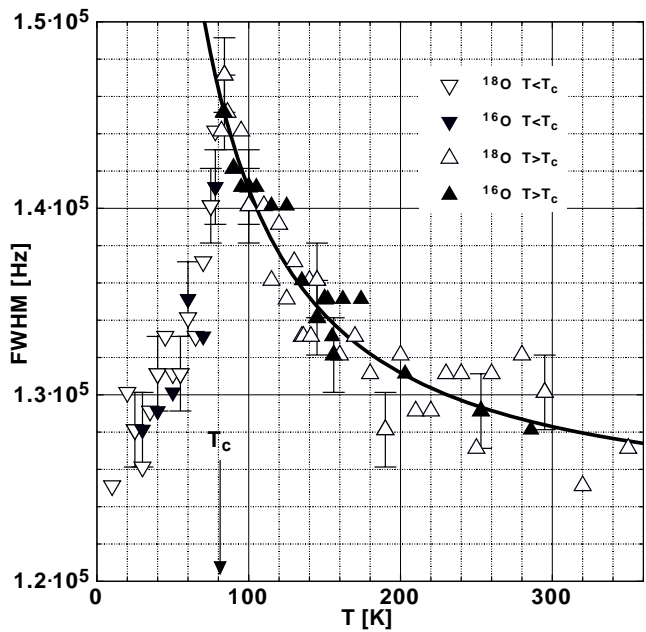} \vskip12pt
 	\caption{Temperature dependence of the linewidth of the plane $^{63}$Cu2 NQR 
 	line in $^{16}$O and $^{18}$O exchanged $\mathrm{YBa_2Cu_4O_8}$. 
 	The solid line is a fit of Eq.~(\ref{lw}) to the linewidth data in the normal 
 	conducting state.}
 	\label{linew}
 \end{figure}

The shape of the Cu2 lines in both oxygen exchanged samples are 
identical and always asymmetric with a tail towards lower frequencies.  
To improve the reproducibility of the analysis we decided to take the 
center of gravity of the upper half of the line as the position of the 
line.  The scatter of the line's position as defined above lies in the 
range of $\pm$ 2~kHz.  Into the same range falls also the scatter of 
the linewidth defined as the full width at the half of the line's 
maximum (FWHM).  Figs.~\ref{freq} and \ref{linew} exhibit the observed 
temperature dependence of the position and the linewidth of the Cu2 
line in $^{16}$O and $^{18}$O $\mathrm{ YBa_2Cu_4O_8}$ samples.  At 
first one clearly notices that the frequency data points of the 
$^{18}$O sample lie beneath the ones from the $^{16}$O sample and that 
the frequency shift of about 15~kHz between the two sets of data is 
roughly constant.  A closer inspection, however, reveals that there is 
also a slight temperature shift between the two temperature 
dependences.  This can be seen best at the temperatures where the 
frequencies have their minima and the temperature dependences their 
largest slope.  These temperatures lie lower for the $^{18}$O sample.  
Further one notices the extraordinarily smooth passage of the 
frequency from the normal to superconducting state.  There is no jump 
in the frequency or even a slight change in its slope at $T_{c}$.  
Also unusual is the rather abrupt saturation of the temperature 
dependence of the frequency below 30~K.  In contrast the linewidth of 
both $^{16}$O and $^{18}$O samples (Fig.~\ref{linew}) show within 
error no difference in the temperature dependence.  The dependence 
itself is surprising and has not been observed in high-$T_{c}$ 
cuprates previously.  Remarkable is the dramatic change of the course 
of the normal state linewidth temperature dependence at $T_{c}$ with 
the subsequent narrowing of the line in the superconducting state such 
that the line towards zero temperature resumes again its high 
temperature normal conducting state linewidth.
\section{DISCUSSION}\label{discussion}
\subsection{NQR frequency}
With help of $\nu_{Q}$ one is able to monitor the electric field 
gradient (EFG) at the nuclear site of interest.  The EFG, a ground 
state property of a solid, depends sensitively on the charge 
distribution in the solid.  Thus knowledge of the EFG and its 
temperature dependence in a high-$T_{c}$ superconductor can contribute 
valuable information concerning the electronic properties of the 
material.  The two naturally occurring copper isotopes $^{63}$Cu and 
$^{65}$Cu have nuclear spin 3/2 and quadrupole moments $^{63}Q$ and 
$^{65}Q$.  In a EFG the Cu nuclei experience NQR at a single frequency 
$\nu^{63,65}_{Q} = (^{63,65}QeV_{zz}/2h) \cdot \sqrt{1+\eta^{2}/3}$ 
where $V_{zz}$ is the major principal axes of the EFG tensor and 
$\eta$ the EFG asymmetry parameter defined as $\eta = 
(V_{xx}-V_{yy})/V_{zz}$.  Here we used the convention: $|V_{xx}| < 
|V_{yy}| < |V_{zz}|$.  To discuss the results a comparison with EFG 
calculations is necessary.  The \emph{ab initio} calculations of the 
EFG \cite{Schwarz,Huesser} are unfortunately not yet precise enough 
for a description of small changes of $\nu_{Q}$.  We use therefore a 
semi-empirical approach to the problem.  We separate the EFG into two 
contributions: (i) the lattice part that refers to the charge 
distribution of the lattice ions surrounding the atom containing the 
nucleus in question and (ii) the valence contribution coming from 
electrons (holes) of the incompletely filled electronic shells of the 
same atom.  The principal components of the EFG tensor $V_{\alpha 
\alpha}$ add as:
$V_{\alpha \alpha} = V_{\alpha \alpha}^{lattice} 
+ V_{\alpha \alpha}^{valence}$, 
where $V_{\alpha \alpha}^{lattice}$ and $V_{\alpha \alpha}^{valence}$ 
denote the lattice and the valence part, respectively. 
Due to the point symmetry at the plane-copper (Cu2) site the principal 
axes of the EFG tensor at this site lie along the crystallographic 
axes with the major principal axis $V_{zz}$ parallel to the crystal 
c-axis \cite{Zimmermann}.  The Cu2 atoms in the CuO$_{2}$ plane act 
closely like Cu$^{2+}$ ions with a hole in the $x^{2}-y^{2}$ orbit.  
Bleaney and coworkers \cite{Abragam} show from electron spin resonance 
studies that a single hole in the d-shell gives rise to an axial EFG 
at the Cu nucleus equivalent to about 70 MHz.  The sign of $V_{\alpha 
\alpha}^{valence}$ from the hole in the $x^{2}-y^{2}$ orbit is 
opposite to the sign of $V_{\alpha \alpha}^{lattice}$ produced mainly 
by the four negatively charged neighbour oxygen atoms.  A rough 
estimate with the so-called point charge model, where the lattice ions 
are approximated by point charges, delivers $V_{\alpha 
\alpha}^{lattice}$ half the size of $V_{\alpha \alpha}^{valence}$.
Therefore in case of Cu2 the sum of the two EFG parts of opposite sign 
turns out to be a difference governed by the larger valence part.  The 
valence contribution to the EFG remains temperature independent (the 
valence does not change) whereas the lattice part diminishes with 
increasing temperature due to the thermal expansion of the lattice.  
Applied to Cu2 this would mean an increase of the summed EFG and thus 
an increase of $\nu_{Q}$ with increasing temperature.  Such a behavior 
is indeed observed for Cu2 in $\mathrm{ YBa_2Cu_4O_8}$, but only for 
temperatures above 200~K.  Below 200~K the behaviour changes 
completely so that towards lower temperatures $\nu_{Q}$ instead 
decreasing starts to grow and finally saturates at 30~K.  Since a high 
resolution neutron powder diffraction study \cite{Kaldis} shows that 
the crystal axes of $\mathrm{YBa_2Cu_4O_8}$ increase smoothly with 
temperature a change of the $\nu_{Q}$ temperature behavior at 200~K 
signals something new, a feature that most likely is connected to the 
electronic system of the CuO$_{2}$ planes.  That the lattice of 
$\mathrm{YBa_2Cu_4O_8}$ indeed behaves in a conventional way one can 
also learn from the temperature dependence of $\nu_{Q}$(Ba), which 
smoothly decreases with temperature from the lowest (5~K) to the 
highest (400~K) temperatures \cite{Lombardi}.  The Ba ion has a filled 
electron shell and therefore experiences only the lattice part of the 
EFG which due to the thermal lattice expansion decreases with 
temperature.  The authors of Ref.  \cite{Lombardi} could fit the 
temperature dependence of $\nu_{Q}$(Ba) quite accurately by a power 
law: $\nu_{Q}(T)=\nu_{Q}(0)(1-AT^{1.43})$.  In an effort to better 
discern the effect of the new electronic feature at low temperatures 
we attempt to extrapolate the normal high-temperature behaviour of 
$\nu_{Q}$(Cu2) to temperatures below 200~K.  In doing so we assume 
that the power law used for $\nu_{Q}$(Ba) holds also for the lattice 
part of Cu2 EFG.  After subtracting this extrapolation from the 
original $\nu_{Q}$(Cu2) data we get a ''rest'' whose temperature 
dependence looks very familiar, namely like the Cu2 Knight shift 
temperature dependence turned on head \cite{Bankay}.  Lacking more 
insight we decide to fit this ''rest'' by an empirical function 
$B\tanh^{2}(\frac{\Delta}{2T})$, that is very reminiscent of the 
empirical spin-pseudogap function used to fit Cu2 Knight shift and 
spin-lattice relaxation data \cite{Raffa,Bankay}.  In the actual 
$\nu_{Q}$(Cu2) vs.  $T$ fit we combine the above two steps and use the 
expression: 
\begin{equation}
	\nu_{Q}(T) = C+AT^{1.43} + 
	B\tanh^{2}(\frac{\Delta}{2T}),
	\label{nuQofT}
\end{equation} 
with the constants $A, B, C$ and $\Delta$ as free fit parameters.  The 
plus sign of the second term in the expression for $\nu_{Q}(T)$ takes 
care of the fact that in case of Cu2, in contrast to Ba, the smaller 
lattice contribution to the EFG gets subtracted from the larger 
valence one.  The obtained fit is quite good (see Fig.~\ref{freq}).  
The corresponding fit parameters are collected in the Table \ref{Table 1}.
The difference in the quantum mechanical zero-point displacement and 
\begin{table}[h]
	\centering
	\caption{The parameters $\Delta$, A, B, and C used in Eq.\ref{nuQofT}) 
	to fit the temperature dependent NQR frequency of plane $^{63}$Cu2 
	in $^{16}$O and $^{18}$O exchanged $\mathrm{YBa_2Cu_4O_8}$.}
	\begin{ruledtabular}
	\begin{tabular}{ccccc}
		  & $\Delta$ (K)& $A$ (MHzK$^{-1.43}$) & $B$ (MHz) & $C$ (MHz)  \\
		\hline
		$^{16}$O & 188.0(1.6) & 2.59(7)$\times 10^{-5}$ & 0.184(2) & 29.642(3)  \\
		$^{18}$O & 180.0(1.6) & 2.69(6)$\times 10^{-5}$ & 0.182(2) & 29.629(2)  \\
	\end{tabular}
	\end{ruledtabular}
	\label{Table 1}
\end{table}  
thermal lattice expansion can explain the nearly constant 15~kHz shift
in the lattice part of the EFG between the $^{16}$O and $^{18}$O 
samples.  The heavier $^{18}$O has a smaller zero-point and thermal 
fluctuation amplitude than $^{16}$O and therefore expands the lattice 
less than $^{16}$O as observed by the x-ray measurement of the lattice 
parameters.  Consequently the lattice part of the EFG in the $^{18}$O 
sample is larger than in the $^{16}$O sample which leads in case of 
Cu2 to a lower $\nu_{Q}$.  Of greater interest is the parameter 
$\Delta$ representing the energy scale of the $\nu_{Q}$ feature.  Its 
value though close to the magnitude of the spin-pseudogap is 
definitely smaller than the later one.  Further, the oxygen isotope 
effect on $\Delta$ is substantial.  The corresponding partial oxygen 
isotope effect coefficient is $\alpha_{\nu_{Q}} = 0.42(11)$ and thus 
much larger than both the spin-pseudogap coefficient $\alpha_{PG} = 
0.061(8)$ and the $T_{c}$ coefficient $\alpha_{T_{c}} = 0.056(12)$ 
\cite{Raffa}.  The EFG does not depend on spin but exclusively on 
charge, however, on all charges, irrespective what state they occupy.  
In that respect the EFG differs from the Knight shift and spin-lattice 
relaxation which experience only the spins of the charge carriers 
whose energies are close to the Fermi level.  Therefore, in case of 
the EFG, at the formation of a pseudogap all the electrons have to be 
considered, not just those close to Fermi surface.  This makes 
conclusions on the basis of an empirical analysis as ours rather 
uncertain.  Nevertheless, the observed large oxygen isotope effect on 
the temperature dependence of $\nu_{Q}$(Cu2) indicates that the 
involved charge feature in the CuO$_{2}$ planes is influenced by the 
coupling of the charge carriers to the lattice as it is the case for 
spin-pseudogap and superconductivity.
\subsection{Linewidth}\label{linewidth}
\begin{figure}[h]
 	\centering
 	\includegraphics[width=\linewidth]{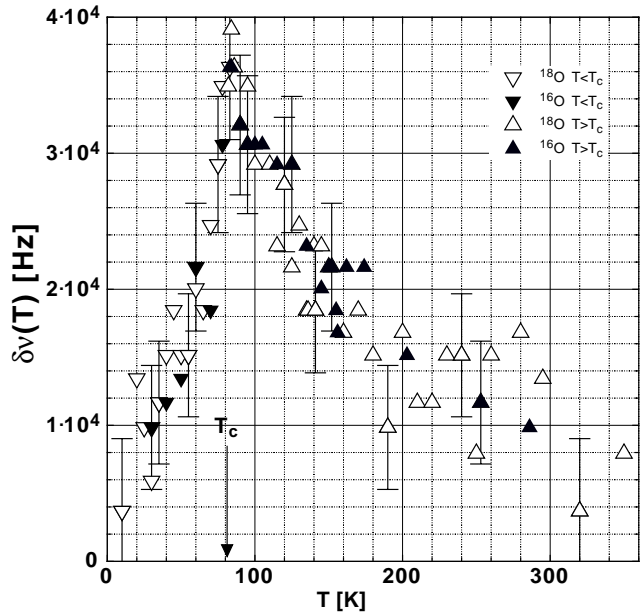} \vskip12pt
 	\caption{The temperature dependent contribution to the 
 	linewidth of the plane $^{63}$Cu2 NQR line in $^{16}$O and $^{18}$O 
 	exchanged $\mathrm{YBa_2Cu_4O_8}$.}
 	\label{Tpart}
 \end{figure}

The linewidths of the Cu2 NQR lines from the two oxygen isotope 
samples (Fig.~\ref{linew}) are equal within error at all temperatures.  
Further, we find that at 350~K the linewidth ratio of the $^{63}$Cu 
and $^{65}$Cu isotope NQR lines is 1.085(7) which within error equals 
to the ratio (1.0806) of the isotopes' nuclear quadrupole moments.  
This allows the conclusion that at 350~K the line is broadened 
predominantly by quadrupolar effects produced by static EFG 
inhomogeneities in the material.  Such inhomogeneities, in principle, 
could be of intrinsic origin generated for instance by a charge 
instability in the electronic system of the plane.  However, this 
possibility most probably can be excluded since a comparison of 
linewidths measured at 350~K in $\mathrm{YBa_2Cu_4O_8}$ samples from 
different batches, shows a variation of these quadrupolar linewidths 
depending on the parameters of material preparation.  Since the 
superconducting properties are not sensitive to the structural 
disorder there is not much incentive to invest for NQR purpose alone 
into the time consuming improvement of the preparation parameters 
necessary to make a structurally perfect material.  Nevertheless, the 
obtained $^{63}$Cu2 line with its 125~kHz linewidth at room 
temperature is to our knowledge the narrowest plane $^{63}$Cu NQR line 
yet observed in high-$T_{c}$ cuprates.  Though this proves the high 
quality of our samples improvements certainly are still possible, 
especially for a stoichiometric material as $\mathrm{YBa_2Cu_4O_8}$ 
where one expects an order of magnitude narrower lines.  
The observed lines are broadened asymmetric having a tail towards 
lower frequencies.  Comparison with the Ba NQR line suggests that this 
asymmetry has to come from a lattice imperfection that produces 
simultaneously an EFG inhomogeneity at Cu2 and Ba sites since the 
shape of Ba NQR line is asymmetric too but with a tail towards higher 
frequencies.  Lattice defects can produce various EFG inhomogeneities.  
The defect we observe, is somewhat special since it can produce 
opposite line asymmetries for Cu2 and Ba NQR lines.  A possible 
candidate that could do so would be for instance a lattice defect that 
locally shrinks the lattice and thus increases the lattice part of EFG 
at both Cu2 and Ba sites.  As already mentioned, any increase of 
lattice EFG pushes Ba and Cu2 NQR frequencies in opposite direction, 
it increases the Ba and decreases the Cu2 one what explains the 
opposite lineshape asymmetries.  The asymmetry of the lines does not 
change with temperature.  Even though asymmetric, the Cu2 line is 
narrow enough to allow a very precise measurement of the broadening of 
the line with decreasing temperature in the normal conducting state 
and what is more important it enables in a high $T_{c}$-cuprate to 
discern how a plane-Cu line gets narrow again below $T_{c}$.  The 
observed change in the temperature behavior at $T_{c}$ is extremely 
sharp with a very rapid decrease of the linewidth at the passage into 
the superconducting state where the line continues to narrow to 
approach at zero temperature its smallest normal state high 
temperature value.  This behaviour, most likely intrinsic in $\mathrm{ 
YBa_2Cu_4O_8}$, has not been observed yet in other high-$T_{c}$ 
cuprates.  It is the excessively broad plane-Cu lines in other 
cuprates that do not allow a similar observation.  Most of cuprates 
are plagued by structural disorder generated by the only possible 
nonstoichiometric doping alternative.  In $\mathrm{YBa_2Cu_4O_8}$ at 
temperatures just above $T_{c}$ where the Cu2 line is broadest the 
ratio of the linewidths of the $^{63}$Cu and $^{65}$Cu isotopes 
decreases from the high temperature value to 1.06(1) which is smaller 
than the ratio of the two isotopes' quadrupole moment and thus signals 
very likely a magnetic component in the line broadening.  In case of 
pure magnetic broadening of the line the ratio of the linewidths of 
the two isotopes would be equal to the quotient of the corresponding 
Cu isotope gyromagnetic ratios $\mathrm{^{63}\gamma/^{65}\gamma 
=0.9335}$.  That a magnetic broadening of the line with decreasing 
temperature indeed takes place we can also infer from the temperature 
dependence of the indirect Gaussian contribution 
$\mathrm{1/T_{2G}^{ind}}$ to the Cu2 spin-spin relaxation rate 
\cite{Stern} that comes from the nuclear magnetization transfer via 
the itinerant electron spin system and which is proportional to the 
static electron spin susceptibility ($\chi(Q_{AF},\omega=0)$) at the 
antiferromagnetic wave vector $Q_{AF}$.  With decreasing temperature 
$\chi(Q_{AF},\omega=0)$ exhibits a Curie like behaviour \cite{Stern}.  
Its contribution to spin-spin relaxation is Curie like and presumably 
the same is true for the linewidth.  To find out the temperature 
dependent contribution to the linewidth we try to deconvolute the 
total linewidth into its temperature independent and dependent parts.  
We find the linewidth at the lowest temperatures to be equal to the 
linewidth at the highest temperature where the magnetic contribution 
due to the Curie like behavior of $\chi(Q_{AF},\omega=0)$ is expected 
to be rather small and the only remaining contribution is then 
quadrupolar as we already noticed before.  Usually a quadrupolar 
contribution that comes from lattice imperfections and has the 
observed large size does not vary much with temperature.  Since we 
find the linewidths at the two temperature extrema equal we assume 
that we have an "underground" quadrupolar contribution that remains 
constant through out the whole temperature range.  In general the 
broadening of a line caused by a combined operation of two distinct 
mechanisms is rather complex.  There are few cases where simple 
relations exist as for instance for two Lorentzian or two Gaussian 
like line broadenings where in the first case the contributions add 
linearly and in the second quadratically.  We do not expect that the 
broadening we study has one of these simple forms.  However, to be 
able to proceed we make an assumption that the constant and the 
temperature dependent part of the linewidth add with equal power $n$.  
We keep $n$ as a free fit parameter with a value that lies somewhere 
between 1 and 2.  We use the linewidth data from the normal conducting 
state as a gauge to find out the temperature dependent part of the 
linewidth in the superconducting state.  For this purpose we fit the 
data gained in the normal conducting state so that we keep the 
underground quadrupolar part of the linewidth constant and allow the 
temperature dependent part to vary Curie like with temperature.  For 
the fit procedure we use the following expression: 
\begin{equation}
	(\delta\nu_{tot}(T))^{n} = (\delta\nu_{quad})^{n} + (C/T)^{n}.
	\label{lw}
\end{equation}	
The fit of the normal conducting 
state data yields:
$\delta\nu_{quad} = 124(1)$ kHz, $C = 3.0(4)$ MHzK, and power 
$n = 1.24(7)$.  
With help of the parameters $\delta\nu_{quad}$ and $n$ we afterwards 
decompose the measured linewidth data to get the wanted temperature 
dependent contribution to the linewidth, $\delta\nu(T)$.  The final 
result is presented in Fig.~\ref{Tpart}.  As one can see the 
temperature dependent part of the linewidth tends to a value close to 
zero when the temperature approaches zero.  The temperature dependent 
part of the linewidth can be of magnetic and of quadrupolar origin.  
Due to the large static electronic susceptibility at $Q_{AF}$ the 
magnetic part of the linewidth comes predominantly from the staggered 
magnetization induced by intrinsic as well as extrinsic magnetic field 
components at $Q_{AF}$. 
From $1/T_{2G}^{ind}$ measurements in the superconducting state 
\cite{Stern} we know that $\chi(Q_{AF},\omega=0)$ decreases only 
15$\%$ of its value at $T_{c}$ when going far into the superconducting 
state.  Therefore, the reduction of the magnetic part of the NQR 
linewidth in the superconducting state has to come from a reduction or 
screening of the extrinsic magnetic fields in the superconducting 
state.  From our measurements it is obvious that with decreasing 
temperature the NQR line broadens smoothly as long as the sample stays 
normal conducting but starts sharply to narrow when the sample turns 
superconducting.  The whole increase in linewidth accumulated from 
350~K down to $T_{c}$ disappears away in the superconducting state.  
At the moment we do not have an adequate explanation of this very 
unusual behavior of the linewidth of the Cu2 NQR line below $T_{c}$.
\section{SUMMARY}\label{summary}
We performed accurate measurements of the temperature dependence of 
the $^{63}$Cu2 NQR line frequency and linewidth in normal and 
superconducting $^{16}$O and $^{18}$O exchanged 
$\mathrm{YBa_2Cu_4O_8}$.  At the transition of $\mathrm{YBa_2Cu_4O_8}$ 
from normal into the superconducting phase we observe that 
$\nu_{Q}$(Cu2) passes $T_{c}$ very smoothly without a discontinuity 
either in value nor in slope.  In contrast to $\nu_{Q}$(Cu2), the 
linewidth exhibits a drastic change of the course at $T_{c}$.  While 
increasing in the normal conducting phase down to $T_{c}$, the 
linewidth starts to decrease sharply in the superconducting phase to 
finally resume towards zero temperature its high-temperature value in 
the normal conducting phase .  Such a behaviour of the narrow 
plane-copper NQR line in $\mathrm{YBa_2Cu_4O_8}$ has not been observed 
yet in other high-$T_{c}$ cuprates.  At the moment we do not have an 
adequate explanation of this extraordinary temperature dependence of 
the Cu2 linewidth.  The frequency $\nu_{Q}$(Cu2) of $^{18}$O exchanged 
$\mathrm{YBa_2Cu_4O_8}$ is always lower than $\nu_{Q}$(Cu2) of 
$^{16}$O exchanged $\mathrm{YBa_2Cu_4O_8}$.  The nearly constant 
frequency shift of 15~kHz can be accounted for by the difference in 
the quantum mechanical zero-point displacement and thermal lattice 
expansion between the $^{16}$O and $^{18}$O samples.  More important, 
there is a well discernible temperature shift between the 
$\nu_{Q}$(Cu2) temperature dependencies of the $^{16}$O and $^{18}$O 
samples.  Lacking a detailed theoretical description we attempt 
empirically to decompose the temperature dependence of $\nu_{Q}$(Cu2) 
into two parts: one part coming from the thermal expansion of the 
lattice and the other produced by the charge redistribution during the 
formation of new electronic structures in the CuO$_{2}$ planes.  From 
the fit of the empirical expression for $\nu_{Q}(T)$ to the 
experimental data we determine for the conjectured formation of new 
electronic structures an energy scale $\Delta = 188.0(1.6)$~K for the 
$^{16}$O and 180.0(1.6)~K for the $^{18}$O exchanged 
$\mathrm{YBa_2Cu_4O_8}$.  The corresponding partial oxygen isotope 
effect coefficient $\alpha_{\nu_{Q}} = 0.42(11)$ is much larger than 
both the spin-pseudogap coefficient $\alpha_{PG} = 0.061(8)$ and the 
$T_{c}$ coefficient $\alpha_{T_{c}} = 0.056(12)$ \cite{Raffa}.

\begin{acknowledgments}
This work was supported in part by the Swiss National Science Foundation.
\end{acknowledgments}


\begin{thebibliography}{}
\bibitem{Raffa} F.~Raffa, T.~Ohno, M.~Mali, J.~Roos, D.~Brinkmann, 
K.~Conder, and M.~Eremin, Phys.~Rev.~Lett. \textbf{81}, 
5912 (1998). 
\relax
\bibitem{Emery}
V.J.~Emery, S.A.~Kivelson, and J.M.~Tranquada, Proc.~Natl.~Acad.~Sci.
USA \textbf{96}, 8814 (1999).
\relax
\bibitem{Zimmermann}
H.~Zimmermann, M.~Mali, D.~Brinkmann, J.~Karpinski, E.~Kaldis, and 
S.~Rusiecki, Physica~C \textbf{159}, 681 (1989). 
\relax
\bibitem{Suter1}
A.~Suter, M.~Mali, J.~Roos, D.~Brinkmann, J.~Karpinski, and E.~Kaldis, 
Phys.~Rev.~B \textbf{56}, 5542 (1997).
\relax
\bibitem{Suter2}
A.~Suter, M.~Mali, J.~Roos, and D.~Brinkmann, Phys.~Rev.~Lett. \textbf{21}, 
4938 (2000).
\relax
\bibitem{Schwarz}
K.~Schwarz, C.~Ambrosch-Draxl, and P.~Blaha, Phys.~Rev.~B \textbf{44}, 5141 (1991).
\relax
\bibitem{Huesser}
P.~H\"{u}sser, H.U.~Suter, E.P.~Stoll, and P.F.~Meier, Phys.~Rev.~B \textbf{62}, 
1567 (2000).
\relax
\bibitem{Abragam}
A.~Abragam and B.~Bleaney, {\em Electron Paramagnetic Resonance of 
transition Ions} (Oxford University Press, New York, 1980).
\relax
\bibitem{Kaldis}
E.~Kaldis, P.~Fischer, A.W.~Hewat, E.A.~Hewat, J.~Karpinski, and S.~
Rusiecki, Physica~C \textbf{159}, 668 (1989).
\relax
\bibitem{Lombardi}
A.~Lombardi, M.~Mali, J.~Roos, and D.~Brinkmann, Physica~C \textbf{267}, 261 (1996).
\relax
\bibitem{Bankay}
M.~Bankay, M.~Mali, J.~Roos, and D.~Brinkmann, Phys.~Rev.~B \textbf{50} 6416 (1994).
\relax
\bibitem{Stern}
R.~Stern, M.~Mali, J.~Roos, and D.~Brinkmann, Phys.~Rev.~B \textbf{51}, 15478 
(1995).
\end{thebibliography}
\end{document}